\pdfoutput=1
\documentclass[11pt]{article}

\pdfoutput=1
\usepackage[]{aaai24}%submission  % DO NOT CHANGE THIS
\usepackage{times}  % DO NOT CHANGE THIS
\usepackage{helvet}  % DO NOT CHANGE THIS
\usepackage{courier}  % DO NOT CHANGE THIS
\usepackage[hyphens]{url}  % DO NOT CHANGE THIS
\usepackage{graphicx} % DO NOT CHANGE THIS
\usepackage{natbib}  % DO NOT CHANGE THIS AND DO NOT ADD ANY OPTIONS TO IT
\usepackage{caption} % DO NOT CHANGE THIS AND DO NOT ADD ANY OPTIONS TO IT
\usepackage{algorithm}
\usepackage{algorithmic}

\usepackage{booktabs}
\usepackage{soul}

\usepackage{xcolor}
\usepackage[most]{tcolorbox}
\newtcolorbox{highlighted}{colback=yellow,coltext=red,breakable}

\usepackage{newfloat}
\usepackage{listings}
\DeclareCaptionStyle{ruled}{labelfont=normalfont,labelsep=colon,strut=off} % DO NOT CHANGE THIS
\floatstyle{ruled}
\newfloat{listing}{tb}{lst}{}
\floatname{listing}{Listing}

\title{How Polarized are Online Conversations about Childhood?}
\author {
    Breanna E. Green\textsuperscript{\rm 1},
    William R. Hobbs\textsuperscript{\rm 2}
}
\affiliations {
    \textsuperscript{\rm 1, 2} Cornell University\\
    \textsuperscript{\rm 1}Information Science\\
    \textsuperscript{\rm 2}Government and Psychology\\
    begreen@infosci.cornell.edu, hobbs@cornell.edu
}

\begin{document}
\maketitle
\begin{abstract}
% The concept of "protecting the children" is a salient political concern in the United States, regardless of party affiliation. Recent focus in this area has impacted law and policy from the overturning of Roe v. Wade to barring gender-affirming care for transgender youth.  With the use of various natural language processing methods on tweets linked to U.S. voters posted between 2019-2022, we sought to analyze these differences in detail. The most important take away of this work is that while both parties do have concerns for protecting children, as evidenced in the similarity of the moral foundation of Care/Harm between groups across time and the results of multivariate regressions, the larger discrepancies in understanding comes in the form of what each party believes is bringing harm to children.
2020 through 2023 were unusually tumultuous years for children in the United States, and children's welfare was prominent in political debate. Theories in moral psychology suggest that political parties would treat concerns for children using different moral frames, and that moral conflict might drive substantial polarization in discussions about children. However, such partisan frames may still differ very little if there is limited underlying disagreement about moral issues and everyday concerns in childhood when not explicitly referencing politics. We evaluate claims of universality and division in moral language using tweets from 2019-2023 linked to U.S. voter records, focusing on \textit{expressed} morality. Our results show that mentions of children by Republicans and Democrats are usually similar, differing no more than mentions by women and men, and tend to contain no large differences in accompanying moral words. To the extent that mentions of children did differ across parties, these differences were constrained to topics polarized well before the pandemic -- and slightly heightened when co-mentioned with `kids' or `children'. These topics reflected a small fraction of conversations about children. Overall, polarization of online discussion around childhood appears to reflect escalated polarization on lines of existing partisan conflicts rather than concerns originating from new concerns about the welfare of children during and after the pandemic.

\end{abstract}

\section{Introduction}
Have online conversations about children become far more polarized since 2019? If so, was increased polarization driven by new concerns about the well-being of children during the pandemic? Or was polarization more narrowly limited to topics that were deeply partisan long before 2020?

In the United States, children's welfare was a leading concern during the COVID-19 pandemic, especially in terms of health (e.g., vaccinations) and negative academic impacts such as severe learning loss  \cite{donnelly2021learning, skar2022learning}. % OLD VERSION: In the United States, Democrats and Republicans tended to disagree about how best to protect children, including how to maintain high-quality education over this period cite \hl{I wonder if we should alter the phrasing here and cite anecdotes / new articles. BG: Maybe remove/cut. How the paragraph begins seems like we might be focusing solely on health + academics.}. There were also a number of politically charged debates on protecting children in 2021 and 2022 that extended beyond direct pandemic-related concerns, including the type of information that should be taught in schools \cite{pewresearchcenter_2022_parentk12school}. The political rhetoric surrounding parental rights extended to current hotbed issues such as abortion, gender-affirming care, and critical race theory (CRT) \cite{cahn2022political, pewresearchcenter_2022_govparentchild}. 
%NEW VERSION:
In 2021 and 2022, Democrats and Republicans engaged in a number of politically charged debates about protecting children that transcended direct pandemic-related concerns, including %the impact of masking mandates \hl{XXcite} and XXnote: WH did you want to keep the masking mandates? i can't remember if I put this here or you; BG: I put that there! I was going to put in an ICWSM cite (the mejova paper) but I moved it to the Moral foundation dictionaries and party differences in moral language use. section when I saw it removed.
the type of information that should be taught in schools \cite{pewresearchcenter_2022_parentk12school}. Further, the political rhetoric surrounding parental rights extended to current hotbed issues such as abortion, gender-affirming care, and critical race theory (CRT) \cite{cahn2022political, pewresearchcenter_2022_govparentchild}. Numerous state and federal level proposed policies either reflected or drove such debates -- including an anti-CRT campaign by Glenn Youngkin in Virginia and Floridian Governor Ron DeSantis's Stop the Wrongs to Our Kids and Employees (W.O.K.E.) Act, as well as the Democratic party's Invest in Child Safety and Raise the Age Acts.

%\hl{Law based essay on 'Parental Rights' as political rhetoric for parental control over current hotbed issues such as abortion,gender-affirming care, and critical race theory } -- \cite{cahn2022political}

%For example, in the 2021 Virginia gubernatorial race, Republican Glenn Youngkin ran a campaign pledging to remove critical race theory (CRT) from school curriculum in order to protect children. Florida Republican Governor Ron DeSantis announced the Stop the Wrongs to Our Kids and Employees (W.O.K.E.) Act which put in place restrictive rules for training and lessons related to diversity, equity and inclusion, or information that makes students feel guilt or psychological distress about the history of the United States. Conversely, the Invest in Child Safety Act was proposed by Democratic U.S. Senators in 2020 to fight against online child exploitation. Both the Raise the Age Act, which sought to raise the purchasing age for semi-automatic rifles from 18 to 21, and the Keep Americans Safe Act, which proposed barring large-capacity magazines, were efforts by Democrats to enforce gun reform.
%In each example, the primary concern shared by Democrats and Republicans is the protection of children, albeit in the ways they respectively believe efforts should take place. 

Any political debate is likely to reflect moral values in one way or another \cite{ryan2014reconsidering}, % WH: mm, I like this too
but debates about children might contain especially prominent moral frames.  % WH: this might be a spot to add more work on the significance of political debates about children
Yet, conversations about children are likely to be more \textit{universally} focused on childhood innocence and the need to protect children \citep[][]{haslam2000essentialist, woodrow1999revisiting}. Everyday parenting concerns often discussed online, such as childcare, screen time, and eating behaviors \cite{thornton2023scoping} might be only rarely polarized. % -- and uncommonly reflecting lifestyle polarization generally \cite{dellapostaWhyLiberalsDrink2015}.

Here, we evaluate the juxtaposition of everyday language and potentially universal moral concerns about protecting children with more divisive partisan frames on social issues. We focus on measuring the extent of polarization in online conversations referencing children. Specifically, we contextualize differences in partisan language with gendered differences and with co-mentions of political topics that were polarized long before the COVID-19 pandemic and later political debates. For this purpose, we use social media posts from 2019 through 2023 by Twitter (later renamed to `\textit{X}') users whose profiles were linked to voter files that include their partisanship and demographics \cite{grinberg2019fake,hughesUsingAdministrativeRecords2021,shugarsPandemicsProtestsPublics2021}.
%Social media platforms, including Twitter, are spaces where political elites share their stances on topics with their constituents, and, importantly, where constituents share their opinions in return. Therefore, using the social media platform Twitter and applying text-as-data and natural language processing methodologies to the accounts of registered U.S. voters, this paper aims to examine if political constituents display distinct differences in their discussions of children and well-being.

In this, we evaluate a claim that we suspect to be true, even during and after the pandemic: that Democrats and Republicans talk about children in the same ways for the vast majority of topics. Clearly illustrating this is nonetheless important. Core similarities might be overlooked \cite{hartmanInterventionsReducePartisan2022} due to strategic use of children's health and welfare in political rhetoric (e.g., to paint an opposition group as more extreme than they really are) and resulting misperceptions of out-party members \cite{wilsonPolarizationContemporaryPolitical2020}. Or due to a general tendency to misattribute disagreements \cite{renDisagreementGetsMistaken2024}. Because of this, highlighting commonalities might itself de-polarize some conversations \cite{ahlerPartiesOurHeads2018,hartmanInterventionsReducePartisan2022,syropoulosEmphasizingSimilaritiesPolitically2023,levendusky2023our}. Relatedly, research focused only on finding differences between parties -- without work that can also demonstrate levels of similarities -- may tend to contribute to such out-party misperceptions. Further, illustrating enduring political differences when they occur can also provide a justification for \textit{acknowledging} (political) differences in conversation when attempting to persuade on politically adjacent topics  \cite{kallaReducingExclusionaryAttitudes2020,hartmanInterventionsReducePartisan2022,eckerPsychologicalDriversMisinformation2022}. This is important in cases where such interpersonal persuasion might improve children's welfare without needing to alter most political identities and beliefs. %For example, because care for children's health and welfare is so universal, it is possible for children to be evoked in political settings as a means to paint an opposition group as more extreme than they really are.}

In our analyses, we evaluate levels of polarization overall and over time using a variety of approaches: moral dictionaries, supervised feature extraction, and (distances in) word embeddings. A major focus in all analyses is not merely \textit{finding} differences but in contextualizing the size of differences. Although we find it possible to mine text data for partisan differences in language use about children, most conversations about children by Democrats and Republicans online are similar. When non-politician Democrats and Republicans do differ slightly, differences seem to more consistently arise out of the expansion of pre-existing political debates -- rather than, as seen during the pandemic, more novel concerns about the welfare of children specifically.% these might be too technical a cite for this purpose since they're mostly about party activists \cite{laymanPartyPolarizationConflict2002,hareConstrainedCitizensIdeological2022}
%Putting this another way, many conversations about children seem to resemble names like ``Stop the Wrongs to Our Kids''. There is universal agreement that children should be protected. Related debates are divisive when they are about the politics of gender, race, and guns more than they are about children.

\section{Background}

% \paragraph{outline}
% \begin{itemize}
%     \item polarization - not sure if will be possible to discuss this well and briefly
%     \item morality
%     \item current events
%     \item children
% \end{itemize}

%In studying levels of political polarization in online conversations about children, we will consider polarization in conversations about children generally as well as polarization in conversations that are more likely to \textit{also} be about politics. 

%In studying levels of moral and political polarization in online conversations about children, we will consider polarization in both general -- potentially apolitical -- and political conversations alike. 

Below, we provide background on partisan variation in moral values and language, and discuss how we might expect the setting of \textit{any} conversation about children online\footnote{Here, specifically on Twitter.} might differ from more focused survey items on political surveys 
% XXnote: \hl{BG to self: return here - connection to surveys} 
\cite[e.g.,][]{barker2006competing} or from the rhetoric of politicians \cite{kraft_klemmensen_2024}. % BG NOTE: need to edit this sentence -->
Drawing on past research, we consider whether to (1) expect new concerns about children during the pandemic to drive political division -- perhaps through competing moral frames applied to all potentially political concerns -- or (2) primarily reflect expanded conflict on already polarized issues such as racism, gender identity, immigration, and gun control.
%To set expectations for the overall analyses, we first consider how likely it is that conversations about children will be polarized whether or not a given conversation is also about politics. 
%We next consider whether conversations more likely to be about politics are more polarized if they either concern new political issues (here, the COVID-19 pandemic) versus more long-standing ones (such as racism, immigration, and gun control). These analyses consider to what extent new concerns about children might drive political division (perhaps through competing moral frames) versus reflecting expanded conflict on already polarized issues. 
Finally, we note conversations about children may differ by time and place. In this, we explain why we might expect Florida to be a most likely case for periods of elevated polarization in conversations about children.

%consider over-time trends in polarization in New York and Florida, assessing whether any spikes in polarized language correspond to pandemic-related or political events. %In the overall versus more explicitly political contrasts, along with over time analyses comparing trends in overall polarization in New York and Florida (\hl{}), provide context for the (often though not always small) levels of increased polarization that we observe in conversations about children.

\subsection{Why online conversations about children might \textit{not} polarize}

% writing out possible text -- will edit and change
% this will need to be cut down/summarized/condensed to allow for a slightly broader set up
%How political are conversations about childhood likely to be? And how much variation in moral framings about childhood can we expect? 
%Under current title/framing, will want to answer why variation in moral framing and not just differences (e.g., in embedding space) by party.

\subsubsection{Morality, politics, and children.}

Theories about morality in politics are useful because they help us abstract away from specific debates and political contexts, linking lines of conflict to more constant differences across political divides. For example, a debate about euthanasia or marijuana can be seen as a manifestation of recurring conflict about purity and sanctity of the human body, and tied to politics through religiosity and conservative ideology \cite{silver2020binding}. %\hl{are my edits here accurate, do they make sense? BG: yes!}

In studying conversations about children, research on morality and politics can be especially informative -- parenting, family, and views toward children play an important role in theories of moral psychology \cite[e.g.,][]{lakoff1996moral,haidt2004intuitive}. Yet, past work might both suggest far-reaching universality in underlying concern for the well-being of children and also extensive polarization in framing and language about how best to protect and provide for children.

An especially prominent theory commonly used in analyses of moral language and politics called moral foundations theory \citep[MFT;][]{haidt2004intuitive, haidt2007morality} argues liberals and conservatives understand morals and base values on five foundations: care/harm, fairness/reciprocity, (in-group) loyalty, authority/respect, and sanctity/purity. Research using MFT has found %\hl{BG: Read the commented out response here if not already read!}%\hl{is this the precise claim of early theoretical work or something that popped out of subsequent empirical work? BG: Subsequent work for sure, once they started associating MFT with partisanship. MFT didn't begin as a politically related theory but rather a cross-cultural theory of differences in ethics/morals. This point would start showing up more in the 2007 Haidt paper cited above and the graham 2009 paper below. WH: Got it! i edited this based on my understanding, do edit if needed} 
 liberal morality to be strongly connected to the care/harm and fairness/reciprocity foundations, while conservative morality is loosely tied to all five foundations but more strongly connected to (in-group) loyalty, authority/respect, and sanctity/purity \cite{graham2009liberals}.

In this, the care/harm foundation is \textit{itself} motivated through a connection to parenthood and hypothesized evolutionary motives to protect vulnerable others, especially children. In text analysis, terms `child' and `children' are regularly included as influential measures of care-related language. Consequently, and despite partisan differences, this theory and related methods based on it might suggest most moral concerns (and, by extension, political concerns) about children would tend to relate to care and protection of them \cite{haidt2007moral}. Thus potentially but not necessarily limiting the amount of division in language about child-related issues. %  (e.g. animals, elderly, children). %On the other, a number studies on using parenting style preferences to measure a preference for more authoritarian government might suggest more general levels of moral and political polarization about childhood \cite{lakoff1996moral, barker2006competing,feinberg2020measuring}.% In considering these \textit{potentially} conflicting strands of work, we will explain that disciplinarian versus nurturing parenting styles can both be used as indicators of preferences for social hierarchy and conformity in survey settings, especially political surveys, without implying that real-world, online conversations about childhood generally will consistently be polarized by political ideology.
%In contrast, more general views toward children tend to be so uniform that , and the care/harm foundation in moral foundations theory is often explicitly tied to parenting. %For example, terms child and children are themselves regularly included as (strong) measures of care-related language, as well as (to a lesser extent) -related language. 
%\hl{The care/harm foundation is derived from an evolutionary connection to parenthood and a need to protect vulnerable others (e.g. animals, elderly, children). This overarching need is universal and potentially apolitical. However, how this need manifests can be political. } 
%Further, although one interpretation of MFT might suggest liberals are more likely to be concerned with the protection of children given their propensity for the care/harm foundation, this idea undermines the traditional family value ideals upheld by conservatives.% Additionally, concerns for the protection of children may be bolstered by religious affiliations, authoritarian parenting style and/or in-group threat, especially among White conservatives. 

\subsubsection{Online social media and political conversations.}

Most prior work on morality in language has
%, rightfully given differing goals, 
focused on more explicitly moral scenarios and contexts. Applications of moral foundations theory to language often focus on politicians' speeches, debates, and social media accounts \cite{deason2012moral, reiter2021studying, hackenburg2023mapping, brisbane2023morality}. A similarly large body of work has studied more typical language related to moral dilemmas in everyday life \cite{kennedy2021moral, nguyen2022mapping, atari2023paucity} and specific public issues (e.g., vaccines, gun control, same-sex marriage, and climate change) \cite{weinzierl2022hesitancy, brady2017emotion}.  

A notable advantage of using data from online social media is that we might expect our findings to differ from previous work due to our focus on very general language. %Our study contrasts with prior work in ways similar to its contrast with closed-ended survey questions about parenting, morality, and political attitudes. 
Though, perhaps surprisingly, \citet{pewresearchcenter_2022_onethirdtweetspolitical} estimates 1/3 of U.S. adults tweets are political, %\footnote{ \url{https://www.pewresearch.org/politics/2022/06/16/politics-on-twitter-one-third-of-tweets-from-u-s-adults-are-political/}} % this is great
the majority of online conversations are \textit{not} about politics. Therefore, we might expect this children \textit{and} explicit politics subset to be relatively small. Conversations might be far more likely to cover, for example, childcare, screen time, and meals \cite{thornton2023scoping}. And so even if there is substantial polarization in political conversations, then their influence on overall levels of polarization in online conversations about children may be limited.

In considering online social media and moral or political conversations, we should note there is an impressive body of machine learning research on classification of moral statements in texts \cite[e.g.,][]{garten2016morality, lin2018acquiring,huang2022learning}. %XX note \hl{more?}.
Our goals here differ from that line of work, as we study how two groups (as linked to voter file demographics) differ \textit{on average}. Precision in our estimates will be largely derived from sample size rather than highly accurate predictions at the document level.

\subsection{Why online conversations about children might (\textit{sometimes}) polarize}

The primary reason we might expect polarization in online conversations is from news reports about the effects of the pandemic, responses to pandemic policies, and waves of new state-level policies related to children and culture war issues.\footnote{See, for example: \url{https://www.washingtonpost.com/education/2022/10/18/education-laws-culture-war/}} There has been seemingly extensive coverage of masking policies, CRT, vaccines, gender identity, remote schooling, and unusually fractious school board meetings. However, such coverage is unlikely to provide a clear answer about the extent of polarization in everyday online conversations. Media coverage is often a poor indicator of the underlying prevalence of a phenomenon \cite{boydstun2013making}, and news reports tend to be focused on particularly novel and news-worthy information. They avoid covering mundane everyday activities, which may or may not be affected by political trends and current events.

Beyond reporting and anecdotes, a number of theories about politics and political polarization have long argued that parenting views reflect, and perhaps drive, political attitudes \cite[e.g.,][]{lakoff1996moral}. There is some empirical work supporting the claim that endorsement of disciplinarian versus nurturing \textit{parenting} styles is related to partisanship as well as a broad range of conservative and liberal political preferences  \cite{barker2006competing,deason2012moral,feinberg2020measuring}.

It is plausible that partisans might then tend to bring similar moral frames to a very wide range of conversations, whether or not a topic has long been explicitly political. For example, polarization in leisure activities and consumer preferences \cite{dellapostaWhyLiberalsDrink2015}, though perhaps through different mechanisms. \citet{janoff2009provide} has argued that although both groups seek to uphold `community' (perhaps a small step from discussing children), conservatives tend to emphasize \textit{protecting the group} and social order, while liberals emphasize \textit{providing for others} and advancing society through change and the promotion of social justice. Although protection of innocence or defending against physical and/or mental harm may be expressed equally between groups -- particularly when talking about children -- it may present in starkly different forms. This may occur at least if a concern can be readily linked to topics commonly associated with social order and justice. 

This said, parenting style scales, many of which are `authoritarianism' scales, use parenting because specific questions related to disciplining and nurturing children are thought to tap into views about social hierarchy, autonomy, and conformity \cite{perezAuthoritarianismBlackWhite2014} or the use of a `nation-as-family' metaphor in how people think about politics \cite{lakoff1996moral} -- not because discussions \textit{about} children are themselves necessarily politically polarized.

\subsubsection{Pandemic and culture war politics: through subsets of conversations and over time in Florida.}

One possible take-away from past work is that we should expect most people to share similar high-level priorities when it comes to children and yet in the context of \textit{discussing} children, we may still see divergent frames evoked by members of different political parties. Similarly, although there is strong evidence of lifestyle polarization \cite{dellapostaWhyLiberalsDrink2015}, such effects seem unlikely to be all encompassing.

That is, although there is likely to be some polarization for at least some topics, we might be less certain how pronounced differing partisan frames might be. Such effects may depend on whether a conversation is more about politics, and associated divisions, than children, and shared experiences and priorities. Further, whether a topic was already polarized prior to the pandemic, and clearly linked to partisan differences in moral values, may influence the extent of polarization on that topic. New topics that may or may not align cleanly on pre-existing divides  might be less readily linked to long-standing divisions for most people, even when there is substantial polarization among party activists and highly political interested people \cite{laymanPartyPolarizationConflict2002}. Thus, in considering general versus domain-specific polarization, we therefore evaluate online conversations about education, pandemic policies, partisanship, and culture war issues (relative to general conversations). % XXnote -- \hl{why these topics} 
These topics were particularly salient between 2020-2023 and offer commensurate co-occurrence with \textbf{children} or \textbf{kids} for comparative study. %XXnote is this poit acceptable to say? I just changed teh timeframe to cover the pandemic categories

%\hl{BG Note: I have some links/articles here in the code editor!!! Are these the type of info you're mentioning related to the impact of Ron Desantis or is this wrong?}
%%% BG NOTES - possible  LINKS FOR FLORIDA importance (and Ron specifically):
%This VOX article is really good! https://www.vox.com/politics/23848897/florida-red-trump-desantis-republican-2024-election
%Ballotpedia info is good also -- https://ballotpedia.org/Party_control_of_Florida_state_government it also uses this data (which might be useful) -->
%NOPE (though interesting)---Gamm, G., & Kousser, T. (2021). Life, literacy, and the pursuit of prosperity: party competition and policy outcomes in 50 states. American Political Science Review, 115(4), 1442-1463.

%Others:
%Using this one! It speaks to him gearing up for he presidential bid, culture wars, and specifically CRT --> DeSantis, G. G. (2024). ‘We believe in education, not indoctrination’: Governor Ron DeSantis, critical race theory, and anti-intellectualism in Florida. Policy Futures in Education, 14782103241226532.

%Zolides, A. (2022). “Don’t Fauci My Florida:” Anti-Fauci Memes as Digital Anti-Intellectualism. Media and Communication, 10(4), 109-117.
%Woody Holton, Chilling Affects: The Far Right Takes Aim at Black History, The American Historical Review, Volume 129, Issue 1, March 2024, Pages 199–216, https://doi-org.proxy.library.cornell.edu/10.1093/ahr/rhae005

We also evaluate the state of Florida as a most likely case for increased salience of pandemic and culture war concerns (relative to the similarly sized by liberal state of New York). 
% XXnote: \hl{why florida} 
We might expect online conversations in Florida, where Governor Ron DeSantis passed a number high salience laws related to politics and children\footnote{\url{https://www.pbs.org/newshour/politics/here-is-a-look-at-the-laws-desantis-has-passed-as-florida-governor-from-abortion-to-guns}.}, potentially as part of a planned run for the U.S. presidency based on culture war messaging \cite{desantis2024we}, to be more polarized than elsewhere. Also, both Florida and New York are distinctive and useful for our analyses because a) their state legislatures and governorships are controlled by a single party (Republicans in Florida and Democrats in New York), making it possible to pass more partisan legislation, b) they are closed primary states, making it possible for us to reliably observe partisan affiliation in voter records, and c) they have large populations, leading to better statistical power\footnote{Nebraska and Wyoming have Republican controlled legislatures and governorships as well as closed primaries, but have far smaller populations than Florida.}. An analysis of Florida versus New York over time may provide additional information on relationships between any existing polarization of conversations and specific events over the 2019 to 2023 period. This can help adjudicate the extent to which polarization might arise relatively directly from the pandemic and children's welfare concerns versus from the extension of existing partisan conflicts.% less clearly related to the pandemic and pandemic-related health and education concerns.

\section{Data}

For our analyses, we use a previously established, large-scale panel dataset which links Twitter data to publicly available voter records, provided through the vendor TargetSmart \citep{grinberg2019fake,hughesUsingAdministrativeRecords2021,shugarsPandemicsProtestsPublics2021}. Approximately 1.5 million Twitter users were matched by unique first and last names to geographical location as of 2018. Due to computational limitations, for the national-level analyses we used a 10\% sample of this data set \footnote{Users whose Twitter user ID's ended with 8, leaving roughly 150k Twitter users from the originally matched data -- many of whom rarely publicly post tweets.}. For the state-level, over-time analyses we retained full samples of Florida and New York rather than a 10\% sample.

%\hl{BG note: it seems like reviewer 4 thinks we ONLY used Florida and NY, rather than US. Tried to organize above}

With this panel, we extracted tweets posted between September 2019 and June 2023 containing terms \textbf{children}, \textbf{kids}, and \textbf{people} (as a comparison). Due to data collection and processing problems in late 2022, we excluded months September through December 2022. These terms are meant to capture direct references of children, and with many fewer misclassifications than terms like \textbf{kid} (e.g., ``I kid'') or \textbf{boys} (often referring to sports or athletes). This avoids methodological artifacts in our word embedding analyses, where a tendency of one party to discuss sports more than the other -- rather than differences in mentions of children -- could drive differences in average word embedding locations. We limited our analyses to users with a provided Democratic or Republican party affiliation (see section Partisanship and gender for more information). Close to 45k of the 150k originally sampled users mentioned \textbf{children} or \textbf{kids} over the study period, of whom approximately 17k were affiliated with the Democratic or Republican party. These 17k users posted 375k tweets mentioning children or kids. Sample sizes for each analysis, including subset analyses, are included in their respective figures.

For extraordinarily active Twitter users to not influence results far more than others, all word embedding analyses and fightin' words %%xxnote: we should keep the name since this isn't juts weighted log odds; BG: got it! WH: see note below -- I'd somehow misunderstood what you meant by weighted log odds
analyses (see Methods section below) consider only one randomly sampled tweet per user. This sampling has the added advantage of reducing computation time, relative to an inversely weighted analysis. It also avoids the need to cluster standard errors by user. Without clustering, we would tend to dramatically overstate the precision of estimates. Analyses with moral dictionaries use user averages rather than tweet sampling. Meaning, we use a user's fraction of tweets with a token matching a moral foundation term.

%\hl{ maybe a table of user and tweet counts}

%We use a previously established, large-scale panel dataset linking Twitter data to publicly available or sourced voter records by matching unique first and last names to geographical location \citep{grinberg2019fake}. Approximately 1.5 million Twitter users were matched with this method as of 2018. From this panel, we extracted tweets posted between January 2019 and July 2022 mentioning terms related to children (i.e. “kids”, “child”, “school age”, etc.) as well as terms relating to family-building ("pregnant") or parenthood ("parent"); (see Appendix Table~\ref{tab:seedwords}). 

\subsubsection{Partisanship and gender.}

To study partisan polarization, we contrasted tweets for Democrats and Republicans. Party affiliations come from voter records for each state, as collected by the voter record vendor. These party affiliations are consistently recorded in states with closed primaries \cite{hughesUsingAdministrativeRecords2021}, meaning for example that only voters who have registered as a Democrat can vote in a Democratic primary. %States without such primaries often do not collect party affiliation. 
To simplify our analyses, we exclude other party affiliations and voters with no indicated party preference\footnote{Either by choice or because a state does not collect the information.}.

We also contrast differences in tweets by gender. Male and female gender has been consistently recorded for all states in the panel, though 1 to 20\% of linked user gender across states is unknown / not provided. This male-female gender comparison provides a benchmark for partisan differences in language use.

%To simplify our political affiliation grouping variable, users were categorized as {\textcolor{blue}{\textbf{Democrat}}}, {\textcolor{red}{\textbf{Republican}}} or {\textcolor{gray}{\textbf{Other}}} based on past voting registration (i.e. Democrat, Republican/Conservative, or Green Party/Libertarian/etc). Moreover, groups in this study should be more broadly understood as Democrats/Left-Leaning/Liberal, Republican/Right-Leaning/Conservative, and Other/Moderate/Unaffiliated. Some states require closed primaries, meaning voters must select a political party when registering to vote. Open primaries dictate that some users in our dataset have no labelled party affiliation. To categorize these users, we relied on their linked partisan score. 

\subsubsection{Text pre-processing.}

We use the quanteda R package \cite{benoitQuantedaPackageQuantitative2018} for text pre-processing prior to our word embedding analysis. With it, we remove punctuation, symbols, numbers, URL's, hashtags, user mentions, and terms that appeared fewer than 5 times in an analysis subset. We also split hyphens and remove separators (meaning, the unicode separator and control categories). For the moral dictionary analyses, we use the tidytext R package \cite{silgeTidytextTextMining2016} with its default settings and without special processing to remove hashtags or usernames.

% \hl{I ended up limiting analyses to just `kids', `children', and `people' (as a comparison) since it appeared that variation in the quality of matching seemed to drive some of the results -- i think this is a more reliable approach}

% Child-related tweets were first selected based on a simple, naive term search via regular expressions (regex). Tweets were then processed by removing punctuation, non-alphanumeric characters, urls, user handles, and hashtags, converting the text to all lowercase, tokenizing, stemming, and filtering out stop words. Other errored matches (e.g. the word \textit{kidding} that matched to \textit{kid} when stemmed) where also removed. We then constructed unigrams, bigrams, and trigrams, and finally extracted exact matches of tokens to the eMFD. A unigram is a single token or single word from a tokenized sentence. 

% Ultimately, we retain 1.5 million child-related tweets posted between January 2019 and July 2022 from approximately 56k individuals.

\section{Methods}

% BG NOTE: Create a diagram/visual here? 

We measure partisan language use differences in multiple, complementary ways: using moral language dictionaries, fightin' words \cite{monroe2008fightin} analysis of moral word use, and word embedding distances. These methods allow us to measure differences on given and previously studied moral dimensions, mine for large differences in language, and measure \textit{any} difference in language use (whether captured by moral dictionaries or not), respectively. 

We contextualize the size of partisan language differences using a gender comparison (for word embedding analyses), variation in differences by associated keywords (education, pandemic, and explicitly partisan keywords), and by contrasting trends in Florida and New York over time. Florida and New York are two states of similar size, with closed party primaries which gives us party affiliation data. Each state had single party control of the state legislature and governorship, therefore their conservative (FL) and liberal (NY) governments had distinct rhetoric and policies related to children over this period.

Each of these analyses contributes to our overarching goal of measuring levels and types of polarization in online discussions about children during the 2019 through 2023 period.

\subsubsection{Moral foundation dictionaries and party differences in moral language use.}

% XXnote: 
% \hl{possibly a brief mention linking this analysis back to the background (why bother with a moral term dictionary -- why not just jump straight to feature extraction and word embeddings)}

% XXnote: 
% \begin{highlighted}
% \textbf{Need to build out or re-think, so boxing this. -- }
% \textbf{WH: This could be an important aside somewhere for ICWSM reviewers -- pointing out that this is not a classification paper. And it should also probably be incorporated somewhere in the data and methods section}

% \hl{BG: I see mention of differences in goal for this work and previous work in section -- \textbf{Online social media and political
% conversations}, should I not repeat here? WH: not sure -- ideally, we'd be able to cite those papers in a more useful + complimentary way (rather than just saying that `we don't do classification like they do, but they're great')}

The prevailing source of moral language identification in text is derived from the original Moral Foundations Dictionary \cite[MFD;][]{graham2009liberals}, and it has been widely used \cite[e.g.,]{johnson2018classification, rezapour2021incorporating, huang2022learning, roy-goldwasser-2023-tale}. %The ability of dictionaries to facilitate word counts or be used as features in classification make such work possible. However, our study on group-based differences in natural language use requires a more in-depth, contextualized understanding of language. Work closer aligned to the current study have used latent semantic analysis \cite{sagi2014measuring}, static word-embedding methods \cite{garten2016morality} or contextual word embeddings \cite{kraft_klemmensen_2024}, but only after using the MFD -- or its extensions \cite{mejova2023authority} -- as a basis. 
Given the paucity of terms in the original MFD (about 320) %other efforts to extend it have been crucial and  enriching \cite{rezapour2019enhancing, araque2020moralstrength, hopp2021extended}. As such, 
we use the Moral Foundations Dictionary 2.0 \citep[MFD2;][]{frimer2019moral} for our analyses of moral word use by party. This dictionary provides  vice and virtue valence lists of words that are `highly prototypic' of a given moral foundation \cite{frimer2019moral}. Words were chosen through expert selection of candidate words, prototypicality analyses using word embeddings, and crowd-sourced validation \cite{frimer2019moral}. 

We model differences by party using linear regression, with robust standard errors, on the fraction of a user's tweets which contained any moral term on a dimension. Party affiliation is the independent variable. We use the fraction of tweets containing moral terms to improve the interpretability of estimates. We also chose the MFD2 over the Extended MFD \cite[eMFD;][]{hopp2021extended}, which relies on term scores rather than prototypical terms, for this reason. Because the goal of our analysis is inference (and the study of averages across parties) rather than user or tweet-level classification, a larger sample size increases the precision of estimates of the group-level averages and comparisons. These analyses were conducted at the user level rather than tweet level. All users in each analysis were weighted equally, in order to avoid giving increased weight to more active users. We further controlled for the pseudo-log, $ln(x+1)$, of a user's average number of tokens per tweet for each given analysis subset to account for higher probability of dictionary matches due to tweet length alone.

From these models we report the coefficients and 95\% confidence intervals for Republican relative to Democrat, with a single moral foundation average as the dependent variable. These estimates are split by vice or virtue terms and by keyword category (see below). We further report sample size, the number of users included in a regression.

\subsubsection{Variation in partisan language use by keyword/topic.}

To better understand variation in partisan language use, we ran analyses on subsets of the data, focusing on those more likely to be related to the pandemic %(and pandemic related outcomes -- here, education) 
or to pre-existing partisan conflicts that appeared to have extended into conversations and political debates about children.

For this, analyses for a given keyword co-occurring in a tweet with \textbf{children} or \textbf{kids} included four categories: education (``teachers'', ``students'', ``schools'', ``books''), pandemic (``vaccine'', ``remote'', ``masks'', ``distancing''), partisanship/ideology (``republicans'', ``liberals'', ``democrats'', ``conservatives), and what we term partisan flashpoints (``trans'', ``racism'', ``migrant'', ``guns''). We chose terms that would be more unequivocal indicators of a topic, that would not be almost solely used by one party or the other (e.g., ``illegals''), and that were relatively frequent in the data. Some more specific phrases, such as ``critical race theory'' or ``CRT'', are very rarely used by most Twitter users. 

%The extended Moral Foundations Dictionary \citep[eMFD;][]{hopp2021extended} is a dictionary based tool constructed from crowd-sourced human coder text annotations. This process extracts the opinions of over 550 raters whose political affiliations and gender balance were sampled to reflect the larger U.S. population. The coders each analyzed twenty texts containing examples of moral conflict in news articles, where they highlighted passages related to each foundation. These highlighted texts were then extracted, analyzed, and assigned foundation probabilities. The final eMFD consists of 3270 words with each term containing a probability across all foundations; a vast extension compared to the original MFD containing slightly over 300 terms.

\subsubsection{Largest term-level party differences in moral foundations.}\label{section:wordassoc}

%% note about being a baysian method
To evaluate the occurrence of moral terms in child-related tweets most associated with each political group, we used a method for determining the weighted log-odds ratios 
of terms being chosen by one group over another \cite{monroe2008fightin}. At a high level, this method identifies terms distinctly used by one group more than another \textit{and} that are relatively frequent. Identified terms are unlikely to randomly occur more times in one group compared to the other. We used the R code (and default prior) provided by its author\footnote{\url{https://burtmonroe.github.io/TextAsDataCourse/Tutorials/TADA-FightinWords.nb.html}}.  

Prior to running this, we removed seed terms \textbf{children} and \textbf{kids} from the analysis. To better understand what moral language is used by each group, we filtered each tweet down to terms that matched the MFD2. %Figure~\ref{fig:moralfw} and Table ~\ref{tab:FW-moral-tbl} show the top terms extracted from FW.

To help interpret these findings, we also ran a bigram version of the same ``fightin' words'' model. This analysis is intended to disambiguate some term meanings and, in some cases, identify possible term-level misclassification of moral content in this dictionary approach. For it, we included both unigrams and bigrams occurring 10 or more times in the analysis data. We then present the top 2 most polarized bigrams along with their matching most polarized unigrams (if any bigrams for a term occurred 10 or more times).

%First, we masked any term that matched the child-related seed words with <MASKTERM>. Doing so reduces the chance for our seed words to overwhelm the log odds with uninformative information, while also preparing the tweets to be scored for eMFD. Next, to better understand what moral language is used by each group, we filtered each tweet down to terms that match the eMFD. Figure~\ref{fig:moralfw} and Table ~\ref{tab:FW-moral-tbl} show the top terms extracted from FW.

%We then use \textbf{eMFDscore}, a Python package developed by \citet{hopp2020scoring} to facilitate the use of the eMFD for calculating moral sentiment and foundation probabilities. 

\subsubsection{Polarization in online conversations about children.}

Current state-of-the-art natural language processing methods have moved beyond dictionary-based methods towards the use of word embeddings (numeric vector representations of words) and other large pretrained language models. More important for this study, recent analyses suggest moral political language may not be adequately captured by dictionaries alone \cite{kraft_klemmensen_2024}. 

To better understand polarization of online conversations about children, %whether or not partisan differences map cleanly onto moral foundations or moral foundation dictionaries, 
we used a modified version of the \textbf{conText} embedding regression method described in \citet{rodriguez2021embedding}. The method takes all the instances where a specific target term (e.g., \textbf{children}) is found in a given corpora, generates GloVe word embeddings from the window of terms surrounding each \textit{context}, and averages across these contexts. For this embedding averaging, we use a window of 6 terms and 200-dimension pre-trained Twitter embeddings \footnote{\url{https://nlp.stanford.edu/projects/glove/}, without the \textit{a la carte} embedding discussed in \citep{rodriguez2021embedding}  -- and so using associations well before our study period.}. Next, it runs linear regressions across all dimensions of the word embeddings. In our case, party (Republican versus Democrat) is the independent variable and a term's averaged embedding of words in its context window on that dimension as the dependent variable. In its original form, the distance between groups is the Euclidean norm of the regression coefficients. We use a corrected version of the original method due to considerable but easy-to-fix bias, as we explain below. Because coefficients from linear regressions (for a two group comparison) represent differences in averages, this calculates the Euclidean distance between the average embedding locations for each group. 

The corrected version of this Euclidean distance estimator removes bias introduced by noise in the estimated regression coefficients \cite{greenMeasuringDistancesHigh2024}. For the squared Euclidean distance, this bias is equal to the sum of the coefficient variances across dimensions,\footnote{As noted in \citet{greenMeasuringDistancesHigh2024}, this method targets an unbiased estimate of the square of a regression coefficient, with population value $\beta^2$. However, we only have an \textit{estimate} of $\beta$, $\hat{\beta}$, and squaring that gives us the expected value: $E[\hat{\beta}^2] = E[\hat{\beta}]^2 + V[\hat{\beta}]$ (from the definition of variance), where $V[\hat{\beta}]$ is the variance of a regression coefficient and $E[\hat{\beta}]^2$ (in expectation) equal to $\beta^2$. Fortunately, standard regression methods provide unbiased estimators for both $\hat{\beta}$ and its variance.} and so we can simply subtract this value to obtain an unbiased estimate of the squared Euclidean distance (Euclidean distance retains a slight bias) \cite{greenMeasuringDistancesHigh2024}. Our estimates use this corrected value for the \textit{squared} Euclidean norm. Note that although squared Euclidean distance is always positive, these estimates are not. Truncating values at 0 would reintroduce bias, as also noted in other very similar corrections \cite[this bias issue has been regularly noted and similarly corrected across fields;][]{weirRoleEcologicalConstraint2012,niliToolboxRepresentationalSimilarity2014,waltherReliabilityDissimilarityMeasures2016}. For these estimates, negative values simply indicate little evidence of embedding differences across groups. 

Last, as noted by \citet{doddBootstrapVarianceSquare2007} and \citet{greenMeasuringDistancesHigh2024}, to our knowledge there is no closed form solution or resampling approach to calculating confidence intervals. Resampling methods provide inaccurate coverage (meaning, a 95\% interval calculated from a bootstrap does not actually cover 95\% of the sampling distribution). Instead, we provide null distributions from permutation tests, which provide accurate p-values for embedding regression \cite{greenMeasuringDistancesHigh2024}. This follows the originally proposed significance testing for embedding regression \cite{rodriguez2021embedding}, which is unaffected by this correction. It is valid in the case of no regression covariates or clustered data -- we sample a single tweet from each user and model using a single independent variable at a time to avoid these complications.

\subsubsection{Over-time analyses.}

%\hl{maybe some of this can be moved up into the background section}

Last, we consider over-time trends in the embedding regression estimates. This matters because we might expect online conversations about childhood to be tied to specific events, especially the pandemic, or to political debates about policies, the 2020 US presidential election, and 2021 handover of power. % XXnote MOVED TO BACKGROUND AS MENTIONED: Further, we might expect online conversations in Florida, where Governor Ron DeSantis introduced a variety of policies related to politics and children, to be more polarized than elsewhere. An analysis of Florida, relatively to a similarly sized but more liberal state (we use New York), over time therefore provide additional information on relationships between any existing polarization and conversations and specific events over the 2020 to 2023 period, partly adjuduciating the extent to which polarization might arise relatively directly from the pandemic versus from the extension of existing conflicts less clearly related to the pandemic and pandemic-related health and education concerns.

In reporting these analyses, we highlight: the start of the pandemic (in March 2020), the murder of George Floyd and the impetus for ensuing racial justice protests (late May 2020), and the inauguration of Joe Biden as US President (in late January 2021). The last two of these events are major political events not related to the pandemic itself and so more likely to be drive shifts in political language about children by partisan conflict extension than pandemic concerns. We also note emergency use authorizations for use of the COVID-19 vaccine in children, starting (for adolescents) in May 2021. See the Data section for why we focus our analysis on Florida and New York.

%%% footnote - using regression for the correction

\section{Results}

\subsubsection{Moral foundations.}

\begin{figure}[!htb]
    \centering
    \includegraphics[width=\linewidth]{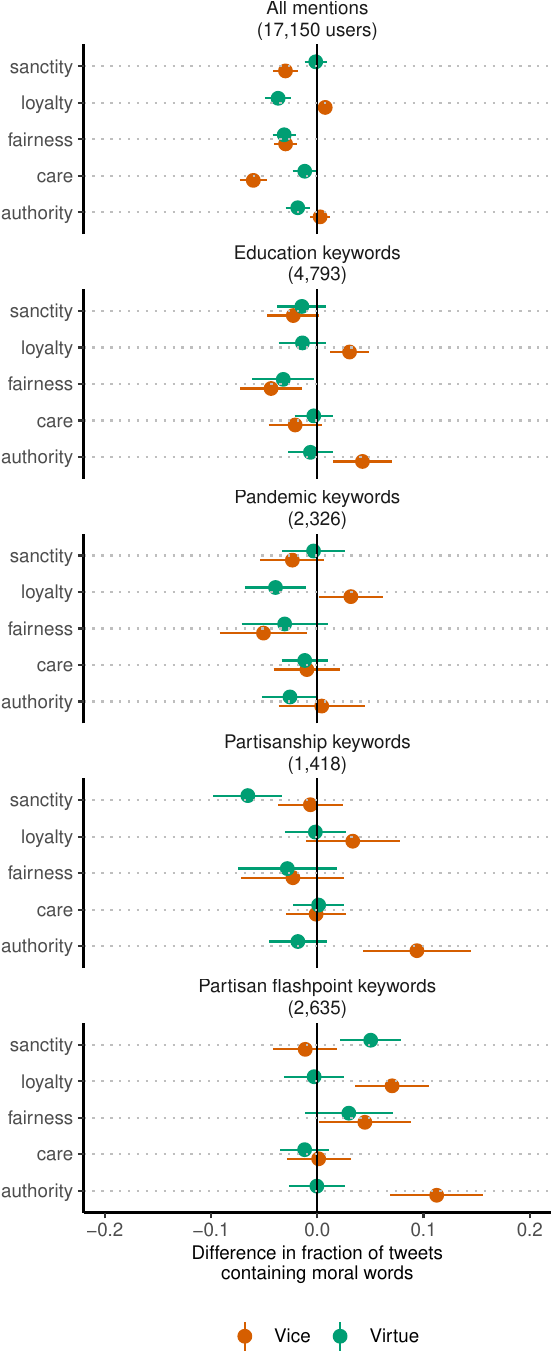}
    \caption{Party moral differences, controlling for differences in tweet length. See text and Figure \ref{fig:emb_by_term} for the lists of keywords by category.}
    \label{fig:moral_by_term}
\end{figure}

In Figure \ref{fig:moral_by_term}, we show differences in the fraction of tweets by party that use a term invoking a particular moral foundation. Similar to past research, we find  Democrats are somewhat more likely to discuss care and fairness than Republicans. We also see Republicans (though less consistently across the vice and virtue terms on a moral foundation) more likely to invoke sanctity, loyalty, and authority for \textit{some} topics. Republicans invoked fairness in relation to partisan flashpoint keywords more so than Democrats.

%\hl{BG: we might not be able to address significance, but it could be worthwhile to mention the Christianity-related words in Republican rhetoric in this context might deal with abortion (if that is true!) or general party related beliefs -- what do you think? (reviewer 2)}

Perhaps surprisingly, many of the estimates do not increase in polarity for keywords more strongly associated with the pandemic or with pre-existing political divisions. In particular, \textit{positive} (i.e., virtue) mentions of a moral foundation are relatively consistent across the subsets. 

At the same time, we do observe larger differences in specific moral foundations. Republicans were substantially more likely to use \textit{negative} (i.e., vice) terms related to loyalty and authority than Democrats for the pandemic and partisan conversation subsets, and especially so for the partisan subsets. We further see some more ambiguous shifts in other negative terms for other moral foundations, where dimensions that saw higher use by Democrats overall saw more equal or slightly greater Republican use in the pandemic and partisan subsets. Nonetheless, these associations did not strongly differ from mentions of \textbf{people}, as we demonstrate in appendix Figure \ref{fig:moral_by_term_people} -- suggesting only very limited partisan moral framing specific to conversations about children.

Because past work suggests that dictionaries may simply miss meaningful polarization \cite{kraft_klemmensen_2024}, this result should be considered in tandem with our word embedding analyses below.

% The overall percentage of each foundation across parties appears to be similarly distributed. Moreover, discussions of harms and children are more prominent than discussions of care and children. 

\subsubsection{Largest term level differences by party.}

Language that differs in moral emphasis often also differs in content. Table \ref{tab:FW-moral-tbl} reports the results of the fightin' words analyses on moral words -- moral terms that tended to be used frequently by one party and far more than by the opposing party. For each term, we list the associated moral foundation, and whether it is categorized as vice or virtue, in the MFD2.  And here, we can see  differences on race (mentioned more by Democrats), %immigration (mentioned more by both parties but in different ways -- as `families' here often referred to separated migrant families, and ``illegals'' referred to migrants), 
health (Democrats), and religion (Republicans). These appear to reflect well-documented differences in issue priorities and religious affiliation by party.\footnote{See, for example: \url{https://www.pewresearch.org/politics/2023/06/21/inflation-health-costs-partisan-cooperation-among-the-nations-top-problems/}, \url{https://www.pewresearch.org/religious-landscape-study/database/party-affiliation/}} In line with past research to some extent, more of Democrats' terms tended to fall under the %the fairness and 
care dimension, and Republicans the sanctity and authority dimensions. %These terms also help explain patterns in Figure \ref{fig:moral_by_term} -- for example, references to ``illegal'' and ``illegals'' can partly explain the greater frequency of authority-vice tweets by Republican in conversations about children.

\begin{table}[!htb]
\centering \resizebox{\columnwidth}{!}{%
%\begin{tabular}{*4c}\toprule
\begin{tabular}{lc}\toprule
Democrat & foundation \\ 
  \hline
fuck (the fuck, fuck you) & sanctity (vice) \\ 
  families (their families, and families) & loyalty (virtue) \\ 
  care (care if, child care) & care (virtue) \\ 
  kill (kill children, monsters kill) & care (vice) \\ 
  health (mental health, the health) & care (virtue) \\ 
  refuse (refuse to, i refuse) & authority (vice) \\ 
  racism (\& racism, racism thread) & fairness (vice) \\ 
  horrific  & sanctity (vice) \\ 
  violence (gun violence) & care (vice) \\ 
  food (food for, food and) & sanctity (virtue) \\ 
  \bottomrule
  Republican & foundation  \\ 
  \hline
jesus  & sanctity (virtue) \\ 
  god (of god, god and) & sanctity (virtue) \\ 
  blessed  & sanctity (virtue) \\ 
  lord (the lord) & sanctity (virtue) \\ 
  father (father of, father left) & authority (virtue) \\ 
  christ  & sanctity (virtue) \\ 
  control (birth control, gun control) & authority (virtue) \\ 
  order (order to, in order) & authority (virtue) \\ 
  cheat  & fairness (vice) \\ 
  sexual  & sanctity (vice) \\ 
% \begin{tabular}{lclc}\toprule
% Democrat & Foundation & Republican & Foundation \\ 
%   \hline
%   fuck & sanctity (vice) & jesus & sanctity (virtue) \\ 
%   families & loyalty (virtue) & god & sanctity (virtue) \\ 
%   care & care (virtue) & blessed & sanctity (virtue) \\ 
%   kill & care (vice) & lord & sanctity (virtue) \\ 
%   health & care (virtue) & father & authority (virtue) \\ 
%   refuse & authority (vice) & christ & sanctity (virtue) \\ 
%   racism & fairness (vice) & control & authority (virtue) \\ 
%   horrific & sanctity (vice) & order & authority (virtue) \\ 
%   violence & care (vice) & cheat & fairness (vice) \\ 
%   food & sanctity (virtue) & sexual & sanctity (vice) \\ 
  \bottomrule
\end{tabular}
}
\caption{In \textbf{children/kids} tweets: top 10 polarized, moral terms and associated foundations in MFD2. Language that differs in moral emphasis also often differs in content. Beyond moral differences, these terms illustrate greater mentions of race and health, as well as negative emotionality (e.g., expletives, violence, injustice), among Democrats on Twitter, and religious language among Republican Twitter users. Terms in parentheses are the most polarized bigrams for a given term, drawn from a separate fightin' words analysis that included both unigrams and bigrams occurring 10 or more times in the analysis data. No bigrams listed indicates that there were no sufficiently frequent bigrams in the analysis data for a given term.} %BG Note: changed 2-gram to bigram for no other reason than I prefer that phrase!
\label{tab:FW-moral-tbl}
\end{table}

\subsubsection{Word embedding distances.}

% To interpret the results of the multivariate regression, we must consider relative distance. In this case, 
The modified embedding regression provides estimates of squared Euclidean distances in a word embedding space between Democrats and Republicans when talking about children on Twitter. If there is little to no difference in word use between groups, the estimates will be closer to zero (or negative, after bias correction), while a larger estimates indicates a larger difference. As noted in the methods section, the \textit{corrected} distance estimator can produce negative values and truncating these values would reintroduce bias in the estimates.

In these models, we compare partisan differences to gender differences, as well as to partisan differences in tweets mentioning \textbf{people} (rather than \textbf{children} or \textbf{kids}). We also analyze overall mentions of children to subsets of mentions that also reference education, pandemic, partisanship, or partisan flashpoints. 

\begin{figure}[!htb]
    \centering
    \includegraphics[width=\linewidth]{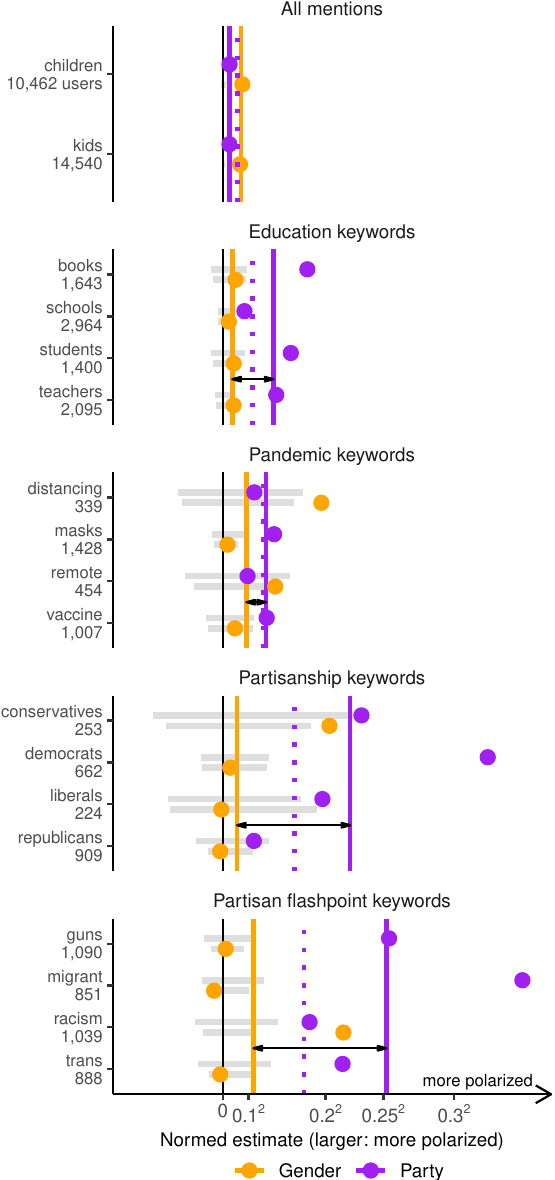}
    \caption{\textit{Embedding Regression by Term.} In this figure, the solid lines indicates the facet (e.g., facet ``Education keywords'') term-frequency weighted average for terms \textbf{children/kids} and the dotted lines facet weighted averages for \textbf{people}. Numbers under each term indicate the number of \textit{users} within the 10\% sample who used that term at least once, and who are included in the term's embedding regression. Gray bars indicate the 95\% confidence intervals for the null distribution of each estimate. These distance estimates can have negative values (see Methods section).}
    \label{fig:emb_by_term}
\end{figure}

Figure \ref{fig:emb_by_term} displays the results. In this figure, points represent the estimated squared Euclidean distance between a given group and the gray lines represent lower and upper 95\% confidence intervals for a null distribution, meaning that points more positive than the gray bars are statistically significant at the 0.05 level. The null distribution is from a permutation test -- as noted in the methods section, bootstrapping does not provide accurate confidence intervals for this distance estimator.

We find overall mentions of children are no more polarized by party than by gender. However, education and pandemic conversations are somewhat more polarized, and conversations that reference partisanship or partisan flashpoints are far more polarized.

% We ran the embedding regression model on sampled instances of selected terms occurring with the term \textbf{child}, then plotted each normed $\hat{\beta}$. Figure \ref{fig:placeholder_} illustrates the term {\textcolor{red}{\textbf{infants}}} is largely understood in a different way between Republicans and Democrats (as shown in {\textcolor{red}{\textbf{red}}}). Compared to other selected terms, {\textcolor{red}{\textbf{infants}}} has the most difference while terms such as {\textcolor{red}{\textbf{guns}}} or {\textcolor{red}{\textbf{mask}}} have the least difference.

% If we take the same terms and incorporate the covariate \textit{sex} (as shown in {\textcolor{yellow}{\textbf{yellow}}}) into our model, we can parse out these differences using our multivariate multiple regression model. Figure \ref{fig:placeholder_} shows differences when controlling for covariates of party affiliation and sex. The coefficient "Male" is the mean difference across sex when controlling for party affiliation. Coefficient "Republican" is the reverse, controlling for gender. In tweets with the word \textbf{child}, only the term {\textcolor{yellow}{\textbf{god}}} is associated with polarization. %(Discuss more -- Use different plot?) 
% Gender is typically a stronger predictor of word use differences. Broadly, this seems consistent with the argument that most discussions of children are not all that morally polarized across party lines. They seem to be polarized for a small subset of conversations.

\subsubsection{Embedding distances over time.}

\begin{figure*}[!htb]
    \centering
    \includegraphics[width=0.8\linewidth]{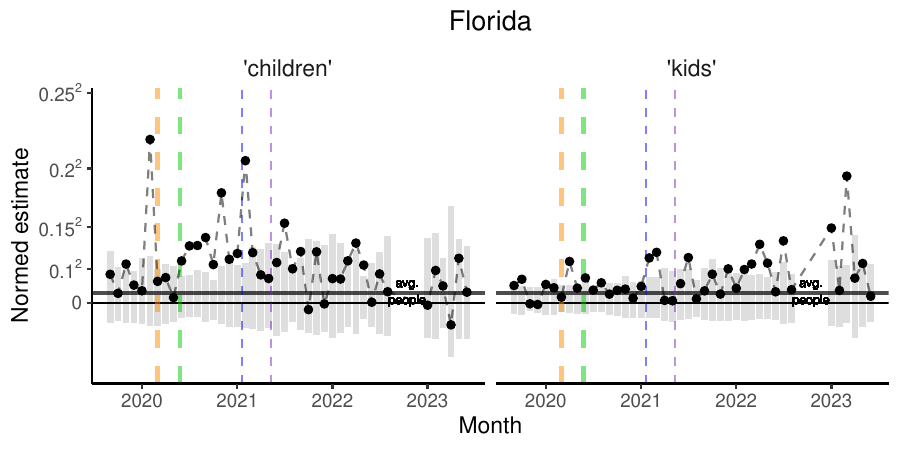}
    \caption{Partisan differences in language use when discussing \textbf{children} and \textbf{kids}: Florida. Vertical lines indicate the start of the COVID-19 pandemic (orange), the murder of George Floyd (green), the inauguration of Joe Biden as US president (blue), and the FDA's emergency use authorization for the COVID-19 vaccine in children aged 12 to 15 in May 2021 (purple). The horizontal black line indicates the average level of partisan difference in language use when mentioning \textbf{people}. Vertical gray bars are 95\% confidence intervals for estimates' monthly null distributions from permutation tests. Months September through December 2022 are missing due to data collection problems during that period.}
    \label{fig:fl_kid}
\end{figure*}

%%% add NY figure
\begin{figure}[!htb]
    \centering
    \includegraphics[width=\columnwidth]{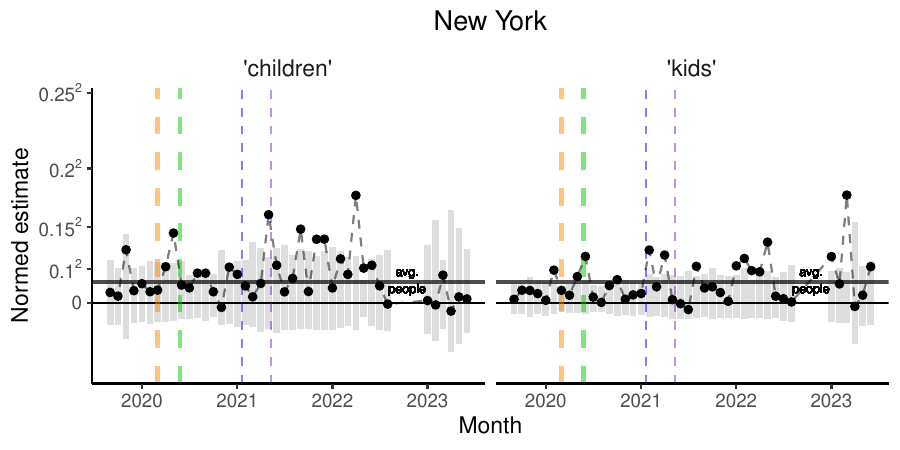}
    \caption{Partisan differences in language use when discussing \textbf{children} and \textbf{kids} -- this figure repeats the analysis in Figure \ref{fig:ny_kid} for the state of New York.}
    \label{fig:ny_kid}
\end{figure}

Finally, we examined differences in the use of \textbf{children} versus \textbf{kids} in a sample of Florida voters. 
%Florida has recently been seen as one of the most prominent states XXnote: \hl{WH: this sort of statement needs to be more concrete (e.g., the Florida governor championed bills to XX, or something about the state government rather than the state as a whole) -- we could remove this sentence here and focus on expanding the explanation for this analysis in the background section. BG: Got it! Most of the support for this is above so might remove from here.} voicing concerns for children across various topics: CRT, LGBTQIA+ rights, schools, vaccines, etc. 
%We sought to explore if there might have been a particular place in time where these conversations and possible differences in language were shown to spike between Democrats and Republicans. 
Figure \ref{fig:fl_kid} illustrates overall partisan polarization in online conversations about children from late 2019 to mid 2023 in Florida. For the term \textbf{children}, we see an increase in polarization in June 2020, just after the murder of George Floyd and several months after the start of the COVID-19 pandemic. Further, we see a large spike in polarization in February 2020 suggesting that conversations about children in Florida may have been polarized before the pandemic as well. For the term \textbf{kids}, we observe a moderate increase in polarization in February 2021, just after the presidential inauguration of Joe Biden.

Patterns in New York State are perhaps less clear. Although we seem to observe a similar pattern for the term \textbf{kids} in New York as in Florida, we do not observe the same increase in polarization in June 2020 for the term \textbf{children}. There is instead an increase in polarization for much of 2021 that, to our knowledge, is not tied to a nationwide political event. Plausibly, it instead aligns with the first emergency authorization for use of a COVID-19 vaccine among adolescents between 12 and 15 in May 2021 \footnote{\url{https://www.fda.gov/news-events/press-announcements/coronavirus-covid-19-update-fda-authorizes-pfizer-biontech-covid-19-vaccine-emergency-use} (the first EUA for adults was December 11, 2020 \url{https://www.fda.gov/news-events/press-announcements/fda-takes-key-action-fight-against-covid-19-issuing-emergency-use-authorization-first-covid-19})}. 

% https://www.fda.gov/emergency-preparedness-and-response/coronavirus-disease-2019-covid-19/covid-19-vaccines

%%% 

%%% 

\section{Discussion}

% BG NOTE: Focus on null results ad implications here!!! Summarize results. My subsubsections here are not final -- just structure for my brain.

This study highlights the limits of moral and political rhetoric in online conversation. We find most Twitter conversations about children among voter file registrants in the United States are not polarized by party, even in the 2020 to 2023 period. 

\subsubsection{Overview of Results.} Conversations related to education and the pandemic were somewhat more polarized, and evoked some divergence in moral frames, while conversations about long-standing partisan divides on race and racism, immigration, gun control, and gender identity were far more polarized. In Florida -- where a number of politically contentious laws related to culture war politics and children have been enacted --  we also observe somewhat greater polarization following political events than in New York. %-- \hl{where Governor Ron DeSantis} %XXnote: \hl{XX BG: what type of point were you thinking here, so I can add to it? Ron's statements, actions, policies, something else? WH: hmm, maybe this isn't necessarily the place to add it (and the background a better spot to add cites like https://www.pbs.org/newshour/politics/here-is-a-look-at-the-laws-desantis-has-passed-as-florida-governor-from-abortion-to-guns). but something about the governor's and state govenrment's rhetoric + policies about children -- just to make clear that Florida should be different (and children + politics potentially more salient) than in New York. this could just hark back to the intro}
%XXnote: I went ahead and took this out. I think it's sufficiently addressed elsewhere. BG: lovely!
%-- 
%than in New York.

%One interpretation of these findings is conversations mentioning children are divisive when they are about the politics of gender, race, immigration, and guns more than they are about childhood or in association with children themselves. %That is, the topics themselves are already polarized, but the additional mention of \textit{children} is used intentionally to bolster the impact.

\subsubsection{Implications.} 
By and large, mentions of children are not polarized by partisanship. That is, when it comes to discussing children in everyday online settings, Democrats and Republicans conversations are mostly similar. In important ways this goes against some recent claims on the increasing politicization of children \cite[e.g.,]{pulciniPoliticizationChildrenImplications2022} that appear to be overly broad. Conversations mentioning children appear to be divisive when they are about the politics of gender, race, immigration, and guns more than they are about childhood. % or in association with children themselves. 
To the extent that polarized issues become more salient in the everyday lives of children, these pre-polarized issues appear to \textit{stay} just as polarized. % in what might otherwise be (mostly) non-partisan conversations.

These findings have a few implications for thinking about and addressing the effects of this \textit{narrow} polarization. First, for the increasing body of research on emphasizing commonalities for reducing the adverse effects of polarization \cite{ahlerPartiesOurHeads2018,hartmanInterventionsReducePartisan2022,syropoulosEmphasizingSimilaritiesPolitically2023,levendusky2023our}, our findings clearly illustrate that there are far more similarities in how partisans talk about children than not. Democrats and Republicans are not coming from different universes on most children's topics, even with polarized partisan messaging during the COVID-19 pandemic. 

Second, however, some issues are political and may stay political, even when they are clearly related to children's welfare (e.g., firearms and firearm safety \cite{mcgoughchild}). Recognizing that conversations on these topics might remain political allows us to take more seriously scientific research on best strategies for effective communication in polarized settings, and that might not be rooted solely on stating evidence or on appeals to authority. Although there is still a lot that is unknown in this research area and there is no foolproof method for avoiding conflict and misunderstanding, strategies might for example eventually include acknowledging and non-judgmentally listening to experiences and political views \cite{kallaReducingExclusionaryAttitudes2020,eckerPsychologicalDriversMisinformation2022,hartmanInterventionsReducePartisan2022}, rather than merely arguing a point. This could be the case even though many people may prefer these conversations to not include politics at all \cite{pulciniPoliticizationChildrenImplications2022,hartmanInterventionsReducePartisan2022}. %Although there is no foolproof method for avoiding conflict and misunderstanding, it is important to recognize the potential relevance of this research (and future advances in it). %More broadly, it highlihts scientific research on productively navigating difficult, political conversations, and best practices tend

%\hl{BG NOTE: musing/don't need to add -- idk if this is a third point or related to one of the points above or below - evoking children in political conversations may or may not contribute to intended impact in the way one would expect;
%we mention the case of guns but should we mention increased polarization for 'migrants'?; are we allowed to show example tweets (I can't remember!)?}

Last, although we do perhaps see somewhat greater polarization when `children' or `kids' are mentioned in specifically flashpoint and partisan topics rather than `people', that difference is smaller than other contrasts (e.g., party versus gender). Further research would be needed to assess the robustness of that increased polarization finding, including across longer spans of time and occurring in other political contexts. It is conceivable that this greater polarization reflects \textit{heightened} attempts to misrepresent and vilify political opponents %\hl{XXcheckcite} I think this works here -- the papers makes the point "political elites may gain more ground by whipping up fury and fear toward opponents than by behaving laudably"
\cite{wilsonPolarizationContemporaryPolitical2020}.

\subsubsection{Limitations.} Although we establish that morally and politically polarized conversations often did not extend into everyday, public, and online conversations about children, a limitation of this study is that studying polarization in conversations about childhood might tell us little about the polarizing \textit{effects} of political rhetoric and pandemic events on attitudes or on \textit{private} and offline conversations. For example, it is plausible that more polarized conversations, especially ones about some types of moral or out-group content \cite{bradyEmotionShapesDiffusion2017,rathjeOutgroupAnimosityDrives2021}, may have been more widely shared and viewed, and so potentially more influential. But even with measures of views, we would still not know whether views of online content translate into attitudes, and there is strong evidence to suggest they often do not \cite{bailAssessingRussianInternet2020,guessHowSocialMedia2023} (though with some exceptions, see for example \cite{mooijman2018moralization}).

Further, we study only the social media platform Twitter. It is possible that we would observe varied effects in general conversations about children on other platforms. In this, we suspect that we would tend to observe those effects on platforms where most conversation is about politics (e.g., on platforms like Gab with a political and largely far-right user base \cite{hobbsAntiMuslimAntiJewishTarget2023}). Of course, we also do not know how far these findings extend into everyday and \textit{private} interactions. There is now some work to rigorously and ethically conduct such studies \cite{reevesScreenomicsFrameworkCapture2021}.

\section*{Ethics Statement}

%All efforts to protect the privacy of individuals in this dataset were taken seriously and with the upmost care. Conducting computational social science work allows for us as researchers to examine societal phenomenon at scale, however that does not overshadow the need for thoughtful ethical considerations. In examining such salient topics as politics and concepts of protections or harms to children, it is important that we take concerted care of the data and privacy of users. 

Conducting computational social science work allows for us as researchers to examine societal phenomenon at scale, however that does not overshadow the need for thoughtful ethical considerations. In examining such salient topics as morality, politics and concepts of protections or harms to children, it is important that we take concerted care of the data and privacy of users. %XXnote: Based on other ICWSM recent papers, a bit of fluff seems encouraged here!
All efforts to minimize risk or harm and protect the privacy of persons in this study were taken. We do not provide user-related information or associated social media text in order to reduce risk of user identification and to avoid violating Twitter terms of service. See \citet{hughesUsingAdministrativeRecords2021} for more details on privacy considerations in the original construction of the Twitter panel from public profiles. All data and figures shown in this paper are displayed in aggregate. This work was approved (as exempt human subjects research) by our university's Institutional Review Board (IRB \#143475 - University affiliation removed for review). We encourage readers of this work to follow the AAAI ethical guidelines \footnote{\url{https://aaai.org/about-aaai/ethics-and-diversity/}} if inspired by the findings.

% \section*{Acknowledgements}

%\newpage

\bibliography{custom,wh_library}
% anthology,
% \bibliographystyle{acl_natbib}

\vfill\eject
%\vspace{10mm}
\Large{\textbf{Appendix}}

\renewcommand{\thefigure}{A\arabic{figure}}

\setcounter{figure}{0}

\begin{figure}[!htb]
    \centering
    \includegraphics[width=0.9\linewidth]{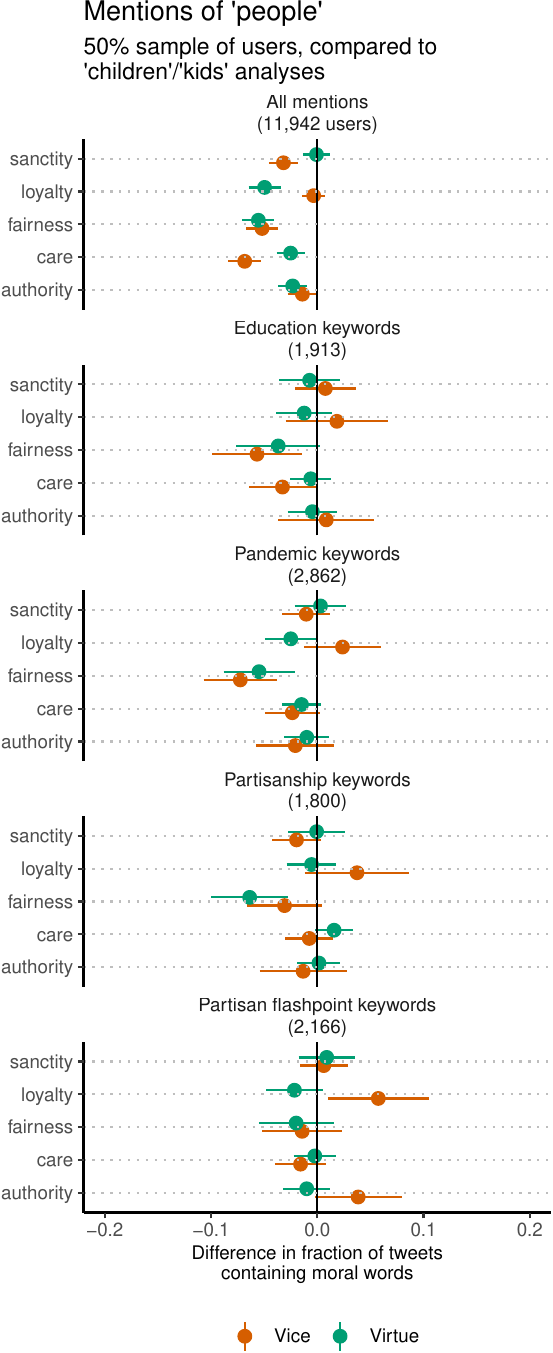}
    \caption{Party moral differences for mentions of \textbf{people}, controlling for differences in tweet length. See text and Figure \ref{fig:emb_by_term} for the lists of keywords by category. Due to computational limitations, we took a further 50\% sample of users for this analysis (multiply the sample sizes here by 2 for a frequency comparison with mentions of \textbf{children/kids}).}
    \label{fig:moral_by_term_people}
\end{figure}

\clearpage
%\appendix

% \clearpage
%^\twocolumn
\vfill\eject

\end{document}